\documentclass[11pt]{article}
\usepackage{amsmath,amssymb,latexsym,bm,graphicx}
\newtheorem{The}{Theorem}
\newtheorem{Pro}[The]{Proposition}

\newtheorem{Exa}{Example}
\newtheorem{remark}{Remark}

\def\qed{\hfill \vrule height 6pt width 6pt depth 0pt}

\usepackage{fancybox}
\usepackage{multirow}

\usepackage{endnotes}
\let\footnote=\endnote
\let\enotesize=\normalsize
\def\notesname{Endnotes}%
\def\makeenmark{$^{\theenmark}$}
\def\enoteformat{\rightskip0pt\leftskip0pt\parindent=1.75em
  \leavevmode\llap{\theenmark.\enskip}}
\def\keywords{\vspace{.5em}
{\textit{Keywords}:\,\relax%
}}

\usepackage{natbib}
 \bibpunct[, ]{(}{)}{,}{a}{}{,}%
 \def\bibfont{\small}%
 \def\bibsep{\smallskipamount}%
 \def\bibhang{24pt}%
 \def\newblock{\ }%

\begin{document}
\title{Exact Simulation Pricing with Gamma Processes and Their Extensions}
\author{Lancelot F. James\\ Department of Information Systems,\\ Business Statistics and Operations Management,\\ Hong Kong University of Science and Technology,\\ Clear Water Bay, Kowloon, Hong Kong.\\\texttt{lancelot@ust.hk} \and Dohyun Kim\\Statistical Research Institute,\\ Seoul National University,\\ Sillimdong, Kwanakgu, Seoul 151-878, Republic of Korea.\\ \texttt{dhkim0211@gmail.com} \and Zhiyuan Zhang\\School of Statistics and Management,\\ Shanghai University of Finance and Economics,\\ No. 777 Guoding Road, Shanghai 200433, China.\\ \texttt{ismtzzy@gmail.com}}
\date{\today}
\maketitle
\newpage

\begin{abstract}
Exact path simulation of the underlying state variable is of great practical importance in simulating prices of financial derivatives or their sensitivities when there are no analytical solutions for their pricing formulae. However, in general, the complex dependence structure inherent in most nontrivial stochastic volatility (SV) models makes exact simulation difficult. In this paper, we shall present a nontrivial SV model that parallels the notable Heston SV model in the sense of admitting exact path simulation as studied by Broadie and Kaya. The instantaneous volatility process of the proposed model is driven by a Gamma process. Extensions to the model including superposition of independent instantaneous volatility processes are studied. Numerical results show that the proposed model outperforms the Heston model and two other L\'evy driven SV models in terms of model fit to the real option data. The ability to exactly simulate some of the path dependent derivative prices is emphasized. Moreover, this is the first instance where an infinite activity volatility process can be exactly applied in such pricing contexts.
\end{abstract}

\newpage

\section{Introduction}\label{sec:Introduction}
Financial models for underlying asset price processes are usually specified as stochastic differential equations (SDE) in the asset pricing context. These include the traditional bivariate diffusion SV models, eg \cite{Heston:1993} model, and the later popular SV models of \cite{BS:2001} (BNS hereafter), where the instantaneous volatility processes are modelled by Non Gaussian Ornstein-Uhlenbeck (OU) processes. The BNS models, due to their flexibility and relative simplicity, have steadily gained in popularity and are now well represented in the literature. See for instance the paper by \cite{CGMY:2003} and monographs of (\cite{CT:2003} and \cite{Schoutens:2003}).

The {\it exact} simulation of such underlying asset price models is of great practical relevance. This is especially true when we can use Monte Carlo simulation to generate an unbiased estimate of the price of a derivative security when there is no analytical solution to its pricing formula. By {\em exact}, we exclude such methods as the first order Euler discretization, which as pointed out for instance in \cite{BK:2006} leads to extra bias and slows down the convergence rate of the estimator from $O(s^{-1/2})$ to $O(s^{-1/3})$ (\cite{DG:1995}), where $s$ stands for the total computational budget. In particular, we mean obtaining {\em exact} sample paths of the state variables. However, in general, the complex dependence structure inherent in most non-trivial SV models makes this task difficult. A notable exception is the work of \cite{BK:2006} who were able to exactly simulate the paths of the Heston model and its jump-diffusion extensions. Therefore, under the Heston model, the square root convergence rate $O(s^{-1/2})$ of the simulation estimator (of eg option price) can be recovered. The Heston model arises by modelling the instantaneous volatility of asset price as a Cox-Ingersoll-Ross (CIR) process whose driving Brownian motion may be correlated with the price process, addressing the possible issue of leverage effect which refers to the phenomenon that volatilities and asset returns are negatively correlated. The simulation techniques used by Broadie and Kaya rely, in part, on the decomposition result of \cite{PY:1982} which are quite specific to the CIR process. However, as pointed out in \cite{BK:2006} the sampling procedure which is based on the numerical inversion of a characteristic function is not without difficulties. In particular \cite{GK:2009} proposes an approximation method that circumvents the use of characteristic function and hence, albeit not technically an exact simulation procedure, alleviate some these difficulties.

A natural question is: are there other price processes, which exhibit a great deal of flexibility, that can be exactly sampled? In this work, we shall demonstrate that a class of models within the BNS framework called BNS OU-Gamma SV model and its extensions is in parallel with the case of \cite{BK:2006}, allowing exact simulation. However, our exact sampling method is entirely different, and we believe much simpler, as it relies only on the simulation of a finite number of independent Uniform[0,1] and Gamma random variables.
Moreover, it is important to mention that, although this is one member of the larger class considered by BNS, we show that, in relation to the Heston model, it is quite comparable, perhaps even better, in terms of model fit to the options market. After all, as we shall see in section \ref{subsec:bivariateDiffu}, the Heston model is also a particular instance of a larger bivariate diffusion class.

First, we show that when the background driving L\'evy process (BDLP) of the instantaneous volatility process, ie a Non Gaussian OU process, is specified by a Gamma process (with scale), the simulation based pricing method can be exactly implemented not only for financial derivatives that depend only on the marginal distribution of terminal asset price (eg European options), both without and with leverage effect, but also for such path dependent derivatives that depend on the joint distribution of asset prices at finitely many time points (eg Forward Start Options). The technical tool we use for the {\it exact} simulation is based on a most recently developed perfect sampling method  of \cite{DJ:2010} termed the Double CFTP (coupling from the past) method. To be specific, the problem of sampling asset prices under the BNS OU SV model is eventually reduced to that of sampling {\it integrated volatilities} (integrals of the instantaneous volatility process over non-overlapped time intervals) possibly along with the corresponding {\it leverage increments} which are BDLP increments due to the component of price process that is used to model the leverage effect. Under the BNS OU-Gamma SV model, these quantities are generalized gamma convolution (GGC) variables and admit {\it independent} decompositions that facilitate exact simulation. See for instance (\cite{BondBook:1992} and \cite{JRY:2008}) on GGC variables. The Double CFTP method is just designed for the exact simulation of this type of variables. This is the first instance where an infinite activity (infinite jumps over any finite time horizon) Non Gaussian OU process can be exactly applied in such pricing contexts. Notice that the exact simulation of the price model is trivial when the BDLP is compound Poisson which is not our interest here. Otherwise, as it is well known, under the BNS framework, an all purpose approach is to apply the infinite series representation method of \cite{Ronsinski:1991}. However, such issues as truncation of the infinite series representation, the slow convergence rate of the series and no explicit expression of the inverse tail mass function (section 5.2.2 of \cite{Schoutens:2003}) in general make the exact simulation difficult for those infinite activity cases. The difficulty arises generally from the nontrivial dependence either between integrated volatility and leverage increment or between integrated volatilities, which as far as we know can only be exactly tackled in the trivial compound case and the OU-Gamma case studied here. Moreover, as we mentioned earlier, in terms of fit of model to the same real option data set, we show in section \ref{subsec:calibration} that the BNS OU-Gamma SV model outperforms both the Heston SV model and the other two BNS OU SV models studied by \cite{NV:2003}.

Second, we provide simulation methods for natural generalizations of the Non Gaussian OU process with Gamma BDLP to the case where the BDLP is a  GGC subordinator, ie increasing L\'evy process without drift that has a GGC marginal distribution, instead of Gamma process, since GGC subordinators include Gamma processes as special cases. Indeed, as shown in section \ref{subsec:sim_GAMMA_GGC}, this extension does provide us more distributional flexibilities. This is due to  the fact that GGC random variables can be represented as scale mixtures of Gamma random variables and hence have distributions that are quite different from the gamma distribution. In fact such random variables may possess heavy tails. We illustrate some simple examples in section 4.

The independence property that we exploit for the BNS OU-Gamma model is not available for the more general BNS OU-GGC models and hence, in general,  we cannot implement an exact simulation procedure. However, in section 4.1 we show that if we modify slightly the framework presented in \cite{BS:2001} it is possible to obtain exact simulation for non-path based options. We refer to this modification as gamma leveraging. In section 4.3 and 5.2,  we discuss highly accurate approximation methods for certain types of path based options.

The rest of this paper is organized as follows: We first give, in section \ref{sec:simulatingPrice}, a general introduction to simulation issues related to pricing financial derivatives under both the bivariate diffusion SV model and the SV model of BNS type.
In section \ref{sec:OUgamma_parallelHeston}, we demonstrate that a BNS type model, ie BNS OU-Gamma SV model, parallels the diffusion type Heston model in the sense of \cite{BK:2006}, admitting exact path simulation. GGC extensions to the OU-Gamma model are made in section \ref{sec:GL_price_processes}. Superposition of independent OU instantaneous volatility processes are also considered in both sections \ref{sec:OUgamma_parallelHeston} and \ref{sec:GL_price_processes}. The Double CFTP exact sampling technique is introduced in section \ref{sec:Sampling}, where two approximation techniques are also discussed. Section \ref{sec:Simulation} and \ref{sec:Numerical_Result} collect all the relevant numerical results while the algorithm of the exact sampler and some of the theoretical details are given in appendices.

\begin{remark}It is well known that due to the presence of Brownian motion, the task of exactly simulating many SV models can be reduced to statements involving the simulation of a joint distribution consisting of an integrated volatility and instantaneous volatility.  This is the case for the Heston model, the general BNS OU class of models, among many others. However what is crucial is to be able to develop an efficient method to exactly sample from the complex joint distributions involving the volatility components. So far the only non-trivial case where this has been achieved is the Heston model. Here we show by quite different sampling techniques that this can also be applied to the BNS OU-Gamma model.
\end{remark}
\begin{remark}
This paper aims at devising exact simulation technique for pricing derivatives whose underlying asset follows a nontrivial BNS OU type SV model that parallels the Heston model. On the other hand, semi-closed expression is available for the price of an European option under Heston model (see \cite{Heston:1993}) and two other BNS OU type models, ie the case where the Non Gaussian OU volatility process has a Gamma marginal distribution (compound case) and the case where the volatility process has an Inverse Gaussian (IG) marginal (see \cite{NV:2003}). These semi-closed price formulae are derived using the inverse Fourier transformation idea that involves corresponding terminal log asset price's characteristic function, whose exponent consists of only elementary complex functions that facilitate the numerical computation, see \cite{Schoutens:2003} pp.87-91 for detailed examples and \cite{CarrMadan:1999} for the computational method Fast Fourier Transformation (FFT). However, under the BNS OU-Gamma/GGC models in this paper, the exponents of characteristic functions consist of nontrivial complex functions (involving even multiple integration of complex functions that need to be evaluated numerically in addition to FFT) that make the implementation of transformation method difficult. See \cite{Zhang:2010} pp.93 for the general characteristic function formula for OU-GGC model.
\end{remark}
\section{Simulating derivative prices under different SV models}\label{sec:simulatingPrice}
We shall consider the problem of pricing a financial derivative with a general discounted payoff function $f(S(t);0\leq t\leq T),$ where $S(t)$ is the underlying asset price process given by certain price model. Based on the fundamental theorem of asset pricing, the no-arbitrage price of the derivative is given by the expected discounted payoff under {\it the risk neutral measure}, ie ${\rm E}[f(S(t);0\leq t\leq T)].$ Whether this expectation can be explicitly evaluated depends on the complexity of both the underlying asset price model $S(t)$ and the payoff function $f(S(t);0\leq t\leq T)$. Notice that in the sequel, we shall always assume that we work with the risk neutral measure (or interchangeably, the equivalent martingale measure) unless otherwise stated.

Under such simple models as Black-Scholes (BS) model, the derivative price is usually explicitly and easily evaluated. However, there is strong empirical evidence of {\it stochastic volatility}, see for instance (\cite{Shephard:1996} and \cite{GHR:1996}), which reflects the deficiency of BS model in describing the underlying asset prices and hence in pricing financial derivatives, since BS model assumes a constant volatility. SV extensions are made to the BS model to circumvent its shortcomings. Under most of those SV models, the evaluation of the above expectation is much more involved and nontrivial anymore. Monte Carlo simulation method could be {\it the} choice whenever the underlying price model or the payoff function is so complicated that analytical evaluation is formidable. The idea is simple, given that we can simulate exactly (if possible) $B$ i.i.d. copies of $\{S(t) : 0\leq t\leq T\},$ ie $\left(\{S^b(t) : 0\leq t\leq T\}\right)_{1\leq b\leq B},$ then the derivative price, ie ${\rm E}[f(S(t);0\leq t\leq T)],$ can be estimated by the following Monte Carlo average
$$
\frac{1}{B}\sum_{b=1}^Bf(S^b(t);0\leq t\leq T).
$$
The estimate converges to the true value as $B\rightarrow\infty$ by the law of large numbers.

In order to make the problem considered tractable enough, throughout, we shall only consider two types of derivatives with the following discounted payoff functions
\begin{quote}
\begin{itemize}
\item[(i)]{\bf (path independent)} $f(S(T));$
    \begin{itemize}
    \item eg European Call Option with discounted payoff ``$e^{-rT}(S(T)-K)^+$'', where $K$ is the strike price and $r$ is the constant risk-free rate.
    \end{itemize}
\item[(ii)]{\bf (path dependent)} $f(S(T_1),S(T_2),\ldots,S(T_m)),$ for $0<T_1<T_2<\ldots<T_m=T.$
    \begin{itemize}
    \item eg Forward Start Option with discounted payoff ``$e^{-rT}(S(T)-kS(T_1))^+$'', where $k$ determines the magnitude of floating strike.
    \end{itemize}
\end{itemize}
\end{quote}
It is readily seen that type (ii) derivatives involve underlying asset prices at finitely many time points while type (i) derivatives depend on only the terminal asset price. As we shall see later, this difference matters a lot when we present simulation issues under the BNS OU SV models.

The rest of this section is devoted to a general introduction of two different types of SV models and their simulation issues related to the above pricing problem.

\subsection{Bivariate diffusion SV model}\label{subsec:bivariateDiffu}
A rather general bivariate diffusion type continuous time models for the underlying asset price process are specified under the risk neutral measure by the following SDE's:
\begin{align}
&d S(t) = (r-q)S(t) dt + \sqrt{v(t)}S(t) dW(t),\notag\\
&dv(t)=\kappa_v(\eta_v-v(t))dt+\sigma_vv(t)^{\alpha_v}dW_1(t),\label{eq:CEVsde}
\end{align}
where $r$ is the risk free rate, $q$ is the dividend yield, $W(t)$ is a standard Brownian motion, $v(t)$ is the underlying instantaneous volatility process that describes the fluctuation of stock price over time, $W_1(t)$ is another standard Brownian motion that may be correlated with $W(t)$ as we shall see later and $\kappa_v$, $\eta_v$, $\sigma_v$ and $\alpha_v$ are positive parameters. The volatility processes given by (\ref{eq:CEVsde}) are called constant elasticity variance (CEV) processes and also used as short-rate model by \cite{KCChan:1992}. When $\alpha_v=1$ and $\alpha_v=1/2$, (\ref{eq:CEVsde}) reduces to the GARCH diffusion process of \cite{Nelson:1990} and the popular CIR process of \cite{CIR:1985} respectively. See also \cite{Meddahi:2002}.

When $\alpha_v=1/2$, we arrive at the popular \citet{Heston:1993} SV model. The following is a typical specification for the price $S(t)$ and volatility $v(t)$ processes under the Heston model:
\begin{eqnarray}
d S(t) &=& (r-q)S(t) dt + \sqrt{v(t)}S(t)( \rho_H dW_1(t) + \sqrt{1-\rho_H^2} dW_2(t))\nonumber\\
dv(t) &=& \kappa_v(\eta_v - v(t))dt + \sigma_v\sqrt{v(t)} dW_1(t). \label{eq:Heston}
\end{eqnarray}
Here $W_1(t)$ and $W_2(t)$ are two independent standard Brownian motions, $\kappa_v, ~\eta_v~{\rm and}~ \sigma_v$ are positive parameters with $2\kappa_v\eta_v\geq\sigma^2_v$ for $v(t)$ never hitting zero (staying positive in other words) and $\rho_H\in[-1,1]$ is the parameter that characterizes the {\it leverage} effect or co-movements between the price and volatility processes. To be precise, $\rho_H$ is the instantaneous constant correlation between the driving Brownian motion of price process, ie $W(t):=\rho_H W_1(t) + (1-\rho_H^2)^{1/2} W_2(t)$, and that of the volatility process, ie $W_1(t)$. Suppose that we are at time $t$ and the future time $T$ is of our concern. Applying stochastic exponential, one can solve for the price of (\ref{eq:Heston}):
\begin{align}
S(T) = S(t)\exp&\{(r-q) (T-t) -\frac{1}{2}\int_t^T v(s)ds + \rho_H \int_t^T \sqrt{v(s)} dW_1(s) \notag\\
&+\sqrt{1-\rho_H^2} \int_t^T \sqrt{v(s)} dW_2(s)\}.
\label{eq:priceTRan}
\end{align}
Simple stochastic integration gives:
\begin{align}
v(T) = v(t) + \kappa_v\eta_v(T-t) - \kappa_v \int_t^T v(s)ds + \sigma_v\int_t^T\sqrt{v(s)} dW_1(s).
\label{eq:volTran}
\end{align}
\citet{BK:2006} proposed an exact simulation method for sampling from the transition distribution of price process under the Heston model (\ref{eq:Heston}), ie $S(T)|S(t),v(t)$, and hence for sampling from the finite dimensional distribution of the price process, ie $(S(T_1),\ldots,S(T_m))|S(t),v(t)$ for $t<T_1<\ldots<T_m$. Denote
$$
\tau(t,T) :=\int_{t}^{T} v(s)ds
$$
that is called {\em integrated volatility} over time interval $[t,T]$. The basic simulation steps of \citet{BK:2006}'s approach can be listed as follows
\begin{quote}

\begin{itemize}
\item[{\em Step 1.}] Sampling $v(T)|v(t)$ that follows a non-central Chi-squared distribution;
\item[{\em Step 2.}] Sampling $\tau(t,T)|v(t),v(T)$;
\item[{\em Step 3.}] $\int_t^T\sqrt{v(s)} dW_1(s)$ is easily recovered through (\ref{eq:volTran});
\item[{\em Step 4.}] Conditional on all the above quantities, $S(T)|S(t)$ is log-normally distributed.
\end{itemize}

\end{quote}
\citet{BK:2006} realizes that the key difficulty arises in Step 2 and proposes a sampling method that is based on inverting the related characteristic function using FFT. This approach is not without any difficulty; see \citet{GK:2009} for more discussions and an alternative simulation procedure under the Heston model.
\begin{remark}
In Step 3, we notice that, due to the square root specification of CIR processes, $\int_t^Tv(s)^{1/2} dW_1(s)$ term that appears as leverage effect in (\ref{eq:priceTRan}) can be recovered through (\ref{eq:volTran}). This fact facilitates the exact path simulation for Heston model. For other more general specifications (\ref{eq:CEVsde}) of $v(t)$ when $\alpha_v\neq1/2,$ exact simulation is much more involved and nontrivial. Notice that CIR process is a particular instance of a larger class (\ref{eq:CEVsde}) of instantaneous volatility models.
\label{rem:heston_simu}
\end{remark}

\subsection{BNS OU SV model}\label{sec:BNS_OU_SV_model}
Continuous time SV models are not limited to be of the bivariate diffusion type. \citet{BS:2001} proposes another type of SV model that is termed BNS OU SV model, where the instantaneous volatility process $v(t)$ is modeled by a Non Gaussian OU process rather than a diffusion process. The change of measure argument of \citet{NV:2003} guarantees that the {\em usual} BNS OU SV model of \citet{BS:2001} with leverage effect can be specified under the risk neutral measure as:
\begin{eqnarray}
dX(t) &=& \left((r-q) -\lambda\kappa(\rho) -\frac{1}{2} v(t)\right)dt + \sqrt{v(t)} dW(t) + \rho d Z(\lambda t)\nonumber\\
dv(t) &=& -\lambda v(t)dt + d Z(\lambda t). \label{eq:BNSOU}
\end{eqnarray}
Here $X(t):=\log(S(t))-\log(S(0))$ is the asset log price, the parameter $\rho\le0$ characterizes the {\em leverage} effect and $\lambda>0$ controls the persistent rate of the instantaneous volatility process $v(t)$. Here $v(t)$ follows a Non Gaussian OU process driven by a subordinator (L\'evy process that has positive increments without drift) $Z(t)$. Notice that
$$
\kappa(\rho) = \int_0^\infty(e^{\rho x}-1)w(x) dx.
$$
Here $w$ is the L\'evy density of the BDLP $Z(t)$.  Models in physical measure and equivalent
martingale measure can be of the same type if the change of measure is {\em structure preserving}. In the sequel, we shall assume the change of measure is structure preserving and work with this risk neutral measure. See \citet{NV:2003} for more details on this. Next, we show how simulation works under the BNS OU SV model. We discuss the path independent and dependent cases and their differences under the BNS framework.

\subsection*{Path independent case:}
First, for the path independent case, the basic problem is the exact sampling from the
distribution of $X(T)|X(t),v(t)$ for some $t<T$. It is easy to see that the distribution of $X(T)-X(t)$ is Normal with mean
\begin{align}
\mu(t,T) = (r-q - \lambda\kappa(\rho))(T-t)-\frac{\tau(t,T)}{2} + \rho\int_t^T dZ(\lambda s)\label{eq:BNSouSVmean}
\end{align}
and variance $\tau(t,T)$ conditional on $X(t),$ $v(t)$, $\tau(t,T)$ and $\int_t^T dZ(\lambda s).$ From \cite{BS:2001} (see also \citet{NV:2003}) we have that:
\begin{eqnarray}
\tau(t,T) &=& \frac{1}{\lambda}\left\{(1- e^{-\lambda (T-t)}) v(t) + \int_t^T(1-e^{-\lambda
(T-s)})dZ(\lambda s)\right\}.\label{eq:BNSouSVvar}
\end{eqnarray}
Hence the problem reduces to how to produce exact i.i.d. samples of the following random pair
\begin{eqnarray}
\left(\int_t^T dZ(\lambda s), \int_t^T (1-e^{-\lambda(T-s)})dZ(\lambda s)\right).\label{eq:Random_Pair_BNS_OU}
\end{eqnarray}
Notice that the first element of (\ref{eq:Random_Pair_BNS_OU}), ie leverage increment, is from the leverage effect while the second element is due to the integrated volatility (\ref{eq:BNSouSVvar}). The corresponding simulation algorithm is
\begin{quote}
\begin{itemize}
\item[{\em Step 1.}] Sampling the random pair (\ref{eq:Random_Pair_BNS_OU});
\item[{\em Step 2.}] Recovering $\mu(t,T)$ and $\tau(t,T)$ through (\ref{eq:BNSouSVmean}) and (\ref{eq:BNSouSVvar});
\item[{\em Step 3.}] Sampling Normal $X(T)-X(t)$ with mean $\mu(t,T)$ and variance $\tau(t,T)$ to get $S(T)=S(t)\exp(X(T)-X(t))$.
\end{itemize}
\end{quote}

\subsection*{Path dependent case:}
Second, for the path dependent case, the basic problem is the exact sampling from the distribution of $(X(T_1),\ldots,X(T_m))|v(t),X(t)$ for $t<T_1<\ldots<T_m$. The difference from the path independent case is realized as follows. Generating $X(T_1)|v(t),X(t)$ can be performed the same way as in the path independent case. However, from $X(T_1)$ to $X(T_2)$ and so on, we need to condition on $v(T_1)$ in order to generate $\tau(T_1,T_2)$ and so on. Denote $T_0:=t,$ $v_0:=v(t),$ $\tau_0:=0,$ $\Delta_i:=T_i-T_{i-1},$ $v_i:=v(T_i),$ and $\tau_i:=\tau(T_{i-1},T_i)$ for $i=1,2,\ldots,m.$ The simulation issue for the path dependent case effectively reduces to how to sample exactly from, in addition to the {\em leverage increment} of (\ref{eq:Random_Pair_BNS_OU}), the series specified by the following two dimensional AR(1) equation (see section 5.4.3 of \cite{BS:2001}):
\begin{align}
\left(
\begin{array} {cll}
\tau_i\\
v_i
\end{array}
\right)&=\left(
\begin{array}{cll}
0 & \frac{1}{\lambda}(1-e^{-\lambda\Delta_i}) \\
0 & e^{-\lambda\Delta_i}
\end{array}%
\right)\left(
\begin{array}{cll}
\tau_{i-1}\\
v_{i-1}
\end{array}
\right)+\left(
\begin{array}{cll}
\frac{1}{\lambda} & -\frac{1}{\lambda}\\
0 & 1
\end{array}
\right)\left(
\begin{array}{cll}
O_{1,i}\\
O_{2,i}
\end{array}\right).\label{eq:TwoDtrans}
\end{align}
Here
\begin{align}
\label{eq:TwoDtransPair}
\left(
\begin{array}{cll}
O_{1,i}\\
O_{2,i}
\end{array}\right)=\left(
\begin{array}{cll}
\int_{T_{i-1}}^{T_i}dZ(\lambda s)\\
\int_{T_{i-1}}^{T_i}e^{-\lambda (T_i-s)}dZ(\lambda s)
\end{array}\right)~{\rm for}~i=1,2,\ldots
\end{align}
and $O_i:=(O_{1,i},O_{2,i})'$s are independent. The simulation algorithm in this case is, for $i=1,2,\ldots,$
\begin{quote}
\begin{itemize}
\item[{\em Step 1.}] Sampling the random pair $O_i$ and noticing that $O_{1,i}$ serves also as the leverage increment, ie the first element in (\ref{eq:Random_Pair_BNS_OU});
\item[{\em Step 2.}] Recovering $\tau_i$ and $v_i$ through (\ref{eq:TwoDtrans}) and $\mu(T_{i-1},T_i)$ through (\ref{eq:BNSouSVmean});
\item[{\em Step 3.}] Sampling Normal $X(T_i)-X(T_{i-1})$ with mean $\mu(T_{i-1},T_i)$ and variance $\tau_i$ to get $S(T_i)=S(T_{i-1})\exp(X(T_i)-X(T_{i-1}))$.
\end{itemize}
\end{quote}

\begin{remark}
\label{rem:pairISSU}
Although the increment $\int_{T_{i-1}}^{T_i}dZ(\lambda s)$ may be the same for both random pairs (\ref{eq:Random_Pair_BNS_OU}) and (\ref{eq:TwoDtransPair}), notice that they come from different parts of the BNS OU SV model. For (\ref{eq:Random_Pair_BNS_OU}), the increment comes from the leverage component of the asset price process, while the increment of (\ref{eq:TwoDtransPair}) comes from the BDLP that drives the instantaneous volatility process. They coincide when the leverage component agrees with the BDLP.
\label{rem:BNSOU_simu}
\end{remark}

In general, sampling {\it exactly} from the random pairs (\ref{eq:Random_Pair_BNS_OU}) and (\ref{eq:TwoDtransPair}) is difficult due to their nontrivial dependence structures, while the case when the BDLP $Z(t)$ is compound Poisson is trivial. Rosinski's (1991) infinite series representation method is usually suggested for sampling the stochastic integrals, ie functionals of the BDLP, that appear in the random pairs; see section 2.5 of \cite{BS:2001}. However, as we mentioned earlier, there is difficulty in implementation of this method as indicated in section 5.2.2 of \cite{Schoutens:2003}, hence the simulation is not {\it exact}. To be specific, besides an infinite series representation, Rosinski's method also involves the evaluation of inverse tail mass function $W^{-1}(x):=\inf\{y>0:W(y)\leq x\}$, where the tail mass function is defined as $W(x):=\int_x^\infty w(y)dy$. In particular, under the proposed OU-GGC model, the L\'evy density of BDLP is of form $w(x)=\theta/x{\rm E}[\exp(-x/R)]$, where $\theta>0$ and the expectation ${\rm E}$ is taken w.r.t. the positive random variable $R$ ($R$ being a constant reduces to the OU-Gamma case). Therefore, the tail mass function under OU-GGC model, being non-standard, is written as
$$
W(x)=\int_x^\infty\frac{\theta}{y}{\rm E}[e^{-y/R}]dy,
$$
whose numerical evaluation is difficult especially when $x$ approaches zero since numerical evaluation of such an integral as $\int_0^11/ydy$ is an ill-posed problem (see section 4.4, chapter 4 of \cite{PTVF:2007} for more details). Moreover, importantly, the inverse of tail mass function is hence not available and numerical inversion is needed presenting additional challenges to implementing Rosinski's method under OU-GGC models. We do not proceed along this course as it is beyond the scope of this paper which after all aims at devising an {\it exact} simulation method.

\section{Exact path simulation of BNS OU-Gamma SV model}\label{sec:OUgamma_parallelHeston}
In this section, we shall present a parallel of \cite{Heston:1993} model within the BNS framework, the BNS OU-Gamma SV model, ie the model (\ref{eq:BNSOU}) with the BDLP $Z(t)$ being specified as a Gamma process with shape parameter $\theta$ and scale parameter $c$. In general, the distributions and the dependence structure of the two components of random pairs (\ref{eq:Random_Pair_BNS_OU}) and (\ref{eq:TwoDtransPair}) are nontrivial. But under the BNS OU-Gamma SV model, the two pairs can be explicitly decomposed as follows:
\begin{eqnarray}
(c\gamma_{\delta},c\gamma_{\delta} M_\delta)\label{eq:Random_Pair_BNS_GAMMA}\\
(c\gamma_{\delta},c\gamma_{\delta} (1-M_\delta)).\label{eq:TwoDtransPair_GAMMA}
\end{eqnarray}
Here $\gamma_\delta$ denotes a ${\rm Gamma}(\delta,1)$ random variable and $M_\delta$ denotes a Dirichlet mean random variable that can be written as the steady solution of the following stochastic equation
$$
M_\delta \stackrel{d}{=}
\beta_{1,\delta}(1-e^{-\lambda U(T-t)}) + (1 - \beta_{1,\delta}) M_\delta,
$$
where $\beta_{1,\delta}$
is a ${\rm Beta}(1,\delta)$ random variable, $U$ is a ${\rm Uniform}(0,1)$ variable, $\delta=\theta \lambda(T-t),$ and all the variables that appear in the right-hand side of the equation are independent of one another. Furthermore, $\gamma_{\delta}$ and  $M_\delta$ are independent, which facilitates the exact simulation. Dirichlet mean random variables are functionals of Dirichlet processes or integrals of functions w.r.t. Dirichlet processes in other words. See \cite{JRY:2008} for a relevant survey and \cite{James:2010} for a recent development. A general theory in this regard is provided in Appendix A. The Double CFTP method of \cite{DJ:2010} can be applied for the {\em exact} sampling of $M_\delta$ in (\ref{eq:Random_Pair_BNS_GAMMA}) and (\ref{eq:TwoDtransPair_GAMMA}), see section
\ref{sec:Sampling} for more details. It remains to obtain $\gamma_\delta$ which is straightforward. Therefore, the path simulation of asset price process, ie sampling prices at finitely many time points, can be implemented exactly, hence the simulation pricing of both path independent and dependent derivatives.

To illustrate how derivative prices can be simulated under the BNS OU-Gamma SV model, we take a look at the path independent case while path dependent case follows similarly. We need to produce an i.i.d. random sample of $S(T)|S(t),v(t)$ to get a Monte Carlo evaluation of the derivative price. For each $b=1,2,\ldots,B$, we generate the pair (\ref{eq:Random_Pair_BNS_GAMMA}), ie $(\gamma^{b}_\delta,\gamma^{b}_\delta M^{b}_\delta)$ and also independently generate $W^{b}$ from a standard Normal distribution. We get the random samples
of $S(T)$ by setting
$$
S^{b}(T) = S(t)\exp(X^{b}(T)-X(t)),
$$
where $S(t)=S(0)\exp(X(t))$,
\begin{align}
X^{b}(T) - X(t) = \left\{r-q +\lambda\theta\log(1-\rho c) \right\}(T-t) -\frac{1}{2} \tau^{b}(t,T)+\sqrt{\tau^{b}(t,T)} W^{b} + \rho c\gamma^{b}_\delta\notag
\end{align}
and
$$
\tau^{b}(t,T) := \frac{1}{\lambda}\left\{(1- e^{-\lambda(T-t)}) v(t) + c \gamma^{b}_\delta
M^{b}_\delta\right\},
$$
by noticing that $\kappa(\rho)=-\theta\log(1-\rho c)$ under the BNS OU-Gamma SV model. As we discussed at the beginning of section \ref{sec:simulatingPrice}, $(1/B)\sum_{b=1}^B f(S^{b}(T))$ would be an unbiased estimate of the true derivative price ${\rm
E}[f(S(T))|S(t),v(t)]$ with discounted payoff function $f(S(T))$.
\begin{remark}
Choosing $Z(t)$ to be a Gamma process for the BNS OU SV model parallels choosing $\alpha_v=1/2$ in (\ref{eq:CEVsde}) under the bivariate diffusion setting. That is, in both cases, in the meantime of tackling the difficult problem of exact simulation from relevant instantaneous/integrated volatilities, one is able to get an exact sample for the leverage effect, see Remark \ref{rem:heston_simu} and \ref{rem:BNSOU_simu}. Table \ref{table:CIR_OU} provides a comparison between the Heston model and the BNS OU-Gamma model in terms of distributions of relevant quantities and their exact sampling methods in the path simulation of underlying asset price process in two time points case.
\end{remark}

\begin{table}[tbhp]\renewcommand{\arraystretch}{1.2}\footnotesize
\centering
    \vspace{0.10in}
    \caption{Comparison between Heston and BNS OU-Gamma models: distributions and sampling methods for related quantities in the exact path simulation.}
    \vspace{0.10in}
    \begin{center}
\begin{tabular}{lccccc}\hline

\multicolumn{2}{c}{}& \multicolumn{2}{c}{Heston} & \multicolumn{2}{c}{BNS OU-Gamma}
\\\cline{3-6}
\multicolumn{2}{c}{}& distribution & sampling method & distribution & sampling method\\\hline\hline
\multirow{2}*{Path-indep.}&$v(T_1)|v_0$ & Noncentral $\chi^2$ &$\chi^2$ with Poisson d.f. &
&\\\cline{2-4} & $\tau_1\mid v_0$ &\multirow{2}*{\citet{BK:2006}}
&\multirow{2}*{FFT}&(\ref{eq:Random_Pair_BNS_GAMMA})\&(\ref{eq:TwoDtransPair_GAMMA})&Double CFTP\\\cline{1-2}
Path-dep.& $\tau_2\mid v(T_1)$&&&&\\
\hline
\end{tabular}
\end{center}
{\it Note.} Involved quantities in the simulation are integrated volatilities, ie $\tau_2 = \tau(T_1,T_2)$ and $\tau_1=\tau(t,T_1)$, and the instantaneous volatilities, ie $v_0=v(t)$ and $v(T_1)$, where $t<T_1<T_2$. Path-dep. is for path dependent, Path-indep. is for path independent and d.f. is for degrees of freedom.
\label{table:CIR_OU}
\end{table}

\subsection{Superposition}\label{subsec:Sup_OU_gamma}
\cite{BS:2001} in section 3 discusses the possibility of generating more flexible dependence structure by the superposition of independent Non Gaussian OU processes, ie weighted sum of independent Non Gaussian OU processes with different persistence rates. \cite{NV:2003} in section 6 also points out the possible research direction of investigating the effect of superposition on option pricing. In the following, we shall show that the previous simulation approach works fine with superposition. That is to say, simulation can be performed exactly for both the path independent and dependent cases under the BNS OU-Gamma SV model with superposition.

One typical log price dynamics of a BNS OU-Gamma SV model with superposition, under the risk neutral measure, may be expressed as:
\begin{eqnarray}
    dX^{s}(t) &=& \left(r-q -\lambda\kappa(\rho) -\frac{v^s(t)}{2}\right)dt + \sqrt{v^s(t)} dW(t) + \rho  dZ(\lambda t),\notag
\end{eqnarray}
where the superscript $s$ of $X^s$ and $v^s$ stands for ``superposition'', the instantaneous volatility is given by the sum of $l$ independent OU processes, ie $v^s(t) =\sum_{j=1}^l v^{(j)}(t)$ for $l\le\infty.$ The $j$th OU process is specified as follows:
$$
dv^{(j)}(t) = -\lambda_j v^{(j)}(t)dt + dZ^{(j)}(\lambda_j t),
$$
where $\lambda_j$'s are chosen such that $\lambda = \sum_{j=1}^l \lambda_j$ and $\left(Z^{(j)}(t)\right)_{1\leq j\leq l}$ are independent Gamma processes with common shape and scale parameters $\theta$ and $c$ such that $Z(\lambda t) = \sum_{j=1}^lZ^{(j)} (\lambda_j t).$ As long as distributions $S^s(T)|S^s(t),{\bf v}(t)$ or $(S^s(T_1),\ldots,S^s(T_m))|S^s(t),{\bf v}(t),$ where ${\bf v}(t) = (v^{(j)}(t); j=1,2,\ldots,l)$ and $S^s$ stands for the corresponding asset price under superposition model, are concerned in pricing financial derivatives, the previous exact simulation strategy readily applies here if we treat each individual $v^{(j)}(t)$ the same way as before. Notice that here the right derivative price is written, for instance, as ${\rm E}[f(S^s(T)) | S^s(t),{\bf v}(t)].$ More technical details can be found in section \ref{subsec:GL_sup}, where we cast the problem in a more general setting.

\section{Extension of OU-Gamma: The OU-GGC Class}\label{sec:GL_price_processes}
A natural extension to the BNS OU-Gamma SV model would be BNS OU-GGC SV model, where we take the BDLP $Z(t)$ of (\ref{eq:BNSOU}) to be a Generalized Gamma Convolution (GGC) subordinator with shape parameter $\theta$ and scale random variable $R,$ ie GGC$(\theta,R).$ Similar extension can also be made to the superposition case.  GGC variables/subordinators can behave quite differently from Gamma ones. We next address this briefly from a density point of view. A Gamma process, ie GGC$(\theta,c)$ ($c$ is a positive constant), evaluated at time 1 is a Gamma$(\theta,1/c)$ random variable with density
$$
\frac{1}{c^\theta\Gamma(\theta)}x^{\theta-1}e^{-x/c}~{\rm for}~x>0.
$$
In the following, we provide two examples of GGC subordinators taken from \cite{James:2010}. We refer to (\cite{JRY:2008} and \cite{James:2010}) for more examples.

\begin{Exa}
For $0<\alpha<1$, let $\mathbb{S}_\alpha$ denote a positive $\alpha$-stable random variable specified by its Laplace transform
$$
{\rm E}[e^{-\omega \mathbb{S}_\alpha}]=e^{-\omega^\alpha}.
$$
Let $\mathbb{S}'_\alpha$ be an independent copy of $\mathbb{S}_\alpha.$ In addition, define
$$
\mathbb{Z}_\alpha=\left(\frac{\mathbb{S}_\alpha}{\mathbb{S}'_\alpha}\right)^\alpha.
$$
The $\Sigma_\alpha(t)$ subordinator in section 4.3 of \cite{James:2010} is a GGC$(1-\alpha,1/\mathbb{G}_\alpha),$ where
$$
\mathbb{G}_\alpha\overset{d}=\frac{\mathbb{Z}_{1-\alpha}^{1/\alpha}}{1+\mathbb{Z}_{1-\alpha}^{1/\alpha}}.
$$
Moreover, $\Sigma_\alpha=\Sigma_\alpha(1)\overset{d}=\frac{\gamma_{1-\alpha}}{U^{1/\alpha}}$ ($U$ is a {\rm Uniform}$[0,1]$ variable) with density function
$$
\frac{\alpha}{\Gamma(1-\alpha)}x^{-\alpha-1}(1-e^{-x}) ~for~x>0.
$$
See \cite{BFRY:2006} for more information on this example.
\end{Exa}

\begin{Exa}
For $c_{et}>0,$ a GGC$(1-\alpha,c_{et}/(c_{et}+\mathbb{G}_\alpha))$, say $\Sigma_{\alpha,c_{et}}(t)$, arises from exponentially tilting GGC$(1-\alpha,1/\mathbb{G}_\alpha)$. The density of $\Sigma_{\alpha,c_{et}}(1)/c_{et}$ is given by:
$$
\frac{\alpha x^{-\alpha-1}e^{-c_{et}x}(1-e^{-x})}{[(c_{et}+1)^\alpha-c_{et}^\alpha]\Gamma(1-\alpha)}~for~x>0.
$$
\end{Exa}
From the above examples, we can see that GGC variables can even have no first moment as in Example 1 in contrast to the Gamma case, for instance, GGC variables $\Sigma_\alpha$ have no first moment for $0<\alpha< 1$. Importantly, our motivation of making the GGC extension is that, as shown in section \ref{subsec:sim_GAMMA_GGC}, OU-GGC models can provide more flexibility in generating different kinds of return distributions.

\subsection{Path independent case: Gamma Leveraging}
As it is usually the case that generalization means loss of certain level of tractability, there is no exception here. Recall that for pricing path independent derivatives, we have to deal with the random pair (\ref{eq:Random_Pair_BNS_OU}), ie $(\int_t^T dZ(\lambda s), \int_t^T (1-e^{-\lambda(T-s)})dZ(\lambda s))$. Nonetheless, unlike the OU-Gamma case, it is difficult to sample exactly the pair (\ref{eq:Random_Pair_BNS_OU}) due to the nontrivial dependence structure of its two components. Recall also Remark \ref{rem:pairISSU} that the necessity of sampling in pair, ie (\ref{eq:Random_Pair_BNS_OU}), is attributable to the usage of the GGC BDLP $Z$ to model the leverage effect. In view of this, we would like to introduce an interesting concept called {\em gamma leveraging}, which means replacing the GGC BDLP $Z(t)$, which appears as the leverage component of the asset log price process in (\ref{eq:BNSOU}), by its {\em extracted} Gamma component. To be specific, the log price process of new {\em gamma leveraged} BNS (GL-BNS) OU-GGC SV model is rewritten as:
\begin{eqnarray}
dX^l(t) &=& \left(r-q -\lambda\kappa_\gamma(\rho) -\frac{1}{2} v(t)\right)dt + \sqrt{v(t)} dW(t) + \rho d \gamma(\lambda t),\label{eq:gammaleverage}
\end{eqnarray}
where the superscript $l$ of $X^l$ refers to ``gamma leveraging'', $\kappa_\gamma(\rho)=-\theta\log(1-\rho)$ and $\gamma(t)$ is the Gamma component of GGC BDLP $Z(t)$, which is made theoretically clear from a Poisson random measure point of view. See Appendix A for details. It then follows that the random pair (\ref{eq:Random_Pair_BNS_OU}) adjusted to the GL-BNS OU-GGC SV model is rewritten as:
\begin{eqnarray}
    \left(\int_t^T d\gamma(\lambda s), \int_t^T (1-e^{-\lambda(T-s)})dZ(\lambda s)\right)
    \stackrel{d}{=} (\gamma_\delta,\gamma_\delta M_\delta),\notag
\end{eqnarray}
where $\delta=\theta\lambda(T-t),$ $M_\delta$ is the steady solution of
$$
M_\delta \stackrel{d}{=} \beta_{1,\delta} R(1-e^{-\lambda U(T-t)}) + (1 - \beta_{1,\delta})M_\delta
$$
and $\gamma_\delta$ and $M_\delta$ are independent. Therefore, exact simulation applies again after the introduction of gamma leveraging. The following Remark \ref{rem:GammaLev} shows that we actually lose no essential economic meaning about the ``leverage effect'' in the gamma leveraged model (\ref{eq:gammaleverage}) comparing to the original BNS OU-GGC model, ie (\ref{eq:BNSOU}) with BDLP $Z(t)$ being given by a GGC subordinator.

\begin{remark}
\label{rem:GammaLev}
Following
\citet{BS:2001}, section 4.3, we can compare the leverage effects, which can be empirically measured by correlation between current return and future volatility, under the BNS OU-GGC model ((\ref{eq:BNSOU}) where $Z(t)$ is GGC) and under the GL-BNS OU-GGC model (\ref{eq:gammaleverage}) by comparing
covariances:
$$
Cov(y_n, y^2_{n+s}) = \frac{\rho {\rm E}[R^2]}{\lambda} (1-e^{-\lambda})^2 e^{-\lambda(s-1)}
$$
and
$$
Cov({\tilde y}_n, {\tilde y}^2_{n+s}) = \frac{{\rm E}[R]}{{\rm E}[R^2]}Cov(y_n; y^2_{n+s}),
$$
where $y_n:=X(n)-X(n-1)$ and ${\tilde y}_n:=X^l(n)-X^l(n-1)$ are the log returns over intervals of length $1$ and $s$ is the length of lag. Hence, in this sense, it is evident that there is not much difference in terms of leverage effects between two models.
\end{remark}

\begin{remark}
We can represent a BNS OU-Gamma model as a GL-BNS OU-GGC model by the reparametrization: $\rho' \leftarrow c\rho,$ where $\rho'$ and $\rho$ are parameters of leverage effect in GL-BNS OU-GGC model and BNS OU-Gamma model respectively. Notice that $R\equiv c$ for an OU-Gamma model.
\end{remark}

\subsection{GL-BNS OU-GGC with superposition}\label{subsec:GL_sup}
A GL-BNS OU-GGC model can also be extended to the case of countable superposition of $l$ independent Non Gaussian OU processes. In this case, the log price model under a risk neutral measure may be expressed as:
\begin{eqnarray}
    dX^{l,s}(t) &=& \left(r-q -\lambda\kappa_\gamma(\rho) -\frac{v^s(t)}{2}\right)dt + \sqrt{v^s(t)} dW(t) + \rho d\gamma(\lambda t),\label{eq:BNS_OU_GAMMA_SUPER}
\end{eqnarray}
where superscript $l,s$ of $X^{l,s}$ refers to both ``gamma leveraging'' and ``superposition'' and $v^s$ is defined similarly as in subsection \ref{subsec:Sup_OU_gamma}. Besides, $Z^{(j)}$'s
are independent GGC subordinators with common shape parameter $\theta$ and corresponding scale random variables $R_j$. Moreover, $\gamma(\lambda t) = \sum_{j=1}^l\gamma^{(j)} (\lambda_j t),$ where $\gamma^{(j)}(t)$ is the extracted Gamma component of its corresponding $j$th GGC BDLP $Z^{(j)}(t).$ Next, we illustrate how path independent derivatives are priced based on this setup.

Let $S^{l,s}(t)$ denote the price corresponding to the model (\ref{eq:BNS_OU_GAMMA_SUPER}) and let
${\bf v}(t) = (v^{(j)}(t); j=1,2,\ldots,l)$. Hence one can simulate the relevant expected discounted payoff function:
\begin{eqnarray*}
{\rm E}[f(S^{l,s}(T)) | S^{l,s}(t),{\bf v}(t)],\label{eq:payoff}
\end{eqnarray*}
if one can exactly sample random variables having the conditional distribution of $S^{l,s}(T)|S^{l,s}(t),{\bf v}(t)$. This can be done as follows. The conditional distribution of $X^{l,s}(T)|X^{l,s}(t),{\bf v}(t)$ is normal with mean
$$
\mu(t,T) = X^{l,s}(t)+(r-q - \lambda\kappa_{\gamma}(\rho))(T-t) -\frac{\tau^s(t,T)}{2}
+ \rho \int_t^Td\gamma(\lambda s)
$$
and variance $\tau^s(t,T)$, given $\int_t^Td\gamma(\lambda s)$ and $\tau^s(t,T).$
Notice that $\tau^s(t,T) = \int_t^T v^s(u)du = \sum_{j=1}^l \tau_j(t,T),$
where each individual integrated volatility can be represented as usual:
$$
\tau_j(t,T) = \frac{1}{\lambda_j}\left\{(1- e^{-\lambda_j(T- t)}) v^{(j)}(t) + \int_t^T
(1-e^{-\lambda_j (T-s)})dZ^{(j)}(\lambda_j s)\right\}.
$$
Hence the simulation problem reduces to how to produce exact samples from the following random pair:
\begin{eqnarray}
    \left(\int_t^Td\gamma(\lambda s), \sum_{j=1}^l\frac{1}{\lambda_j}\int_t^T
(1-e^{-\lambda_j (T-s)})dZ^{(j)}(\lambda_j s)\right).\label{eq:Random_Pair_GL_Super}
\end{eqnarray}
Proposition \ref{prop:sup} in the following characterizes the distribution of the pair (\ref{eq:Random_Pair_GL_Super}) that enables one to use the Double CFTP for exact sampling.

\begin{Pro}
Consider specifications for the pair in
(\ref{eq:Random_Pair_GL_Super}) where $Z^{(j)}$s are independent GGC$(\theta,R_j)$ subordinators. Here $(R_1,\ldots,R_l)$ are independent variables with distributions
$(F_{R_1},\ldots,F_{R_l}).$ Set $\lambda_j = \lambda p_j$ with $\sum_{j=1}^l p_j=1.$ Then, for
$\delta = \theta\lambda(T-t)$, the joint distribution of
(\ref{eq:Random_Pair_GL_Super}) is equivalent in distribution to the vector:
$$
(\gamma_{\delta}, \gamma_{\delta} M_{\delta}),
$$
where $\gamma_{\delta}$ is independent of $M_{\delta}$, which satisfies
$$
M_{\delta}\stackrel{d}{=} \beta_{{\delta},1}M_{\delta} +
(1-\beta_{{\delta},1})O_L,
$$ where $O_L \stackrel{d}{=} R_L
[\lambda_L^{-1} (1-\exp(-\lambda_L U(T-t)))]$, $L$ is a discrete random variable on
$\{1,2,\ldots,l\}$ with $P(L=j) = p_j$, independent of $(R_1,\ldots,R_l)$. If the
$R_j\stackrel{d}{=}R$ are identically distributed, then $O_L \stackrel{d}{=} R [\lambda_L^{-1}
(1-\exp(-\lambda_LU(T-t)))]$.
\label{prop:sup}
\end{Pro}

\textit{Proof}. By straightforward calculations, the L\'evy exponent
of
$$
\sum_{j=1}^l \frac{1}{\lambda_j}\int_t^T (1-e^{-\lambda_j(T-s)})dZ^{(j)}(\lambda_j s)
$$
can be expressed as
$$
\theta\lambda(T-t) \sum_{j=1}^l p_j {\rm E}[\log(1+\omega R_j
\lambda_j^{-1}(1-e^{-\lambda_jU(T-t)}))]
$$
and this is the same as $\theta\lambda(T-t) {\rm E}[\log(1+\omega O_L)]$, concluding the result.
\qed

\subsection{Path dependent case: Approximations}\label{subsec:PathDepAppIntro}
As we remarked in section \ref{sec:BNS_OU_SV_model}, see Remark \ref{rem:pairISSU}, in order to simulate path dependent derivative prices, one needs to be able to sample exactly, in addition to the {\em leverage increment} of (\ref{eq:Random_Pair_BNS_OU}), the random pair (\ref{eq:TwoDtransPair}), ie
$$
(\int_{T_{i-1}}^{T_i}dZ(\lambda s),\int_{T_{i-1}}^{T_i}e^{-\lambda(T_i-s)}dZ(\lambda s)),
$$
where the increment of $\int_{T_{i-1}}^{T_i}dZ(\lambda s)$ comes from the BDLP $Z(t)$, which drives the instantaneous volatility process $v(t)$, rather than the leverage component of the price process. Hence, under the GL-BNS OU-GGC SV model, the Double CFTP method can not be directly applied to the exact path simulation because of the nontrivial dependence between the two elements in (\ref{eq:TwoDtransPair}). Further, the gamma leveraging introduced before cannot help tackling this difficulty in the path dependent case. However, as we shall demonstrate in the following section, several approximation methods perform surprisingly well in handling the above situation.

\section{Sampling GGC random variables}\label{sec:Sampling}
As we have already seen that in both the OU-Gamma and OU-GGC models simulation effectively involves the decomposition of a stochastic integral like $\int_t^Tk(s)dZ(\lambda s)$, where the BDLP $Z(t)$ is a Gamma process or GGC subordinator. The possible choices in this paper for $k(s)$ are: $1,$ $\exp(-\lambda(T-s))$ and $1-\exp(-\lambda(T-s)).$ Under both models, this integral is a GGC random variable with decomposition like $\gamma_\delta M_\delta,$ where $\gamma_\delta$ is a ${\rm Gamma}(\delta,1)$ variable and $M_\delta$ is a Dirichlet mean random variable that is, as we mentioned before, the steady solution of a stochastic equation of form:
\begin{eqnarray}
M_\delta \stackrel{d}{=} \beta_{1,\delta}Y + (1 - \beta_{1,\delta}) M_\delta,\label{eq:MVY2}
\end{eqnarray}
where $Y$ is a positive scale random variable. Corresponding to the above three choices of $k(s),$ $Y$ is $R,$ $R\exp(-\lambda U(T-t))$ and $R(1-\exp(-\lambda U(T-t)))$ respectively, where $R$ is the scale variable of GGC BDLP $Z(t)$ and $R\equiv c$ reduces to the Gamma BDLP case. Moreover, $\gamma_\delta$ and $M_\delta$ are independent. The next section discusses the {\em exact} sampling method for Dirichlet mean variable $M_\delta.$ In subsection \ref{subsec:PathDep_APP} we discuss several approximate sampling methods and how the nontrivial dependence structure of the pair (\ref{eq:TwoDtransPair}) is handled by those methods under the GL-BNS OU-GGC models as mentioned earlier in subsection \ref{subsec:PathDepAppIntro}. The precision of approximation is confirmed by a comparison with the exact Double CFTP method in section \ref{sec:Simulation}.

\subsection{Exact method: The Double CFTP}\label{subsec:CFTP}
The first type of algorithm is an {\em exact} sampling method by \cite{DJ:2010} which is called the double coupling from the past (Double CFTP). \citet{PW:1998a,PW:1998b} proposed coupling from the past (CFTP) method that is an exact sampling algorithm for the steady state distribution of a Markov chain. In the CFTP, the past time is identified by the first time
that every possible state is coalesced into a single state at the current time, and that single state is exactly from the steady state distribution of the Markov Chain, see \cite{DJ:2010} for a review of CFTP. By using a doubling trick, \cite{DJ:2010} proposed the Double CFTP algorithm which finds the coalescence time and hence generate a sample from the steady state distribution of Markov Chain determined by the equation (\ref{eq:MVY2}). If $Y$ is a bounded random variable and $\delta\le1$ then the Double CFTP can be directly applied to sample exactly from the Dirichlet mean random variable $M_\delta$. When $\delta>1$, we cannot directly adopt the Double CFTP method. Nonetheless, let $0<\delta_j\le 1,~j=1,\ldots,l $ such that $\sum_{j=1}^l\delta_j = \delta.$ Then $M_\delta$ can be represented as:
$$
M_\delta \stackrel{d}{=} \sum_{j=1}^l \frac{G_j}{G}M_{\delta_j},
$$
where $G_j$'s are independent ${\rm Gamma}(\delta_j,1)$ random variables, $G=\sum_{j=1}^l G_j,$ $M_{\delta_j}$'s are independent Dirichlet mean variables with shape parameters $\delta_j$ and common scale random variable $Y$ and, moreover, $G_j$'s are independent of $M_{\delta_j}$'s. In summary, we can use the Double CFTP method for exact simulation as long as $Y$ is a bounded random variable. Appendix B summarizes the algorithm.
\begin{remark}
From Figure \ref{fig:ES} in section 6, we notice that the function $f$ which maps $\delta\in(0,1]$ to the expected stack size is convex. By the property of convex function, we have $f(\delta/l)\leq1/l\sum_{j=1}^lf(\delta_j)$ for $\delta>1$ and any decomposition such that $\delta_j\in(0,1]$. Therefore, it is easy to see that the best decomposition for any non-integer $\delta>1$ is: $\delta_i=\delta/l,$ for $i=1,\ldots,l,$ where $l=\lfloor\delta\rfloor+1.$
\end{remark}

\subsection{Approximation methods}\label{subsec:PathDep_APP}
The second type of algorithms are approximation methods based on truncation with both fixed and random number of components. A Dirichlet mean random variable defined through (\ref{eq:MVY2}) can be represented as:
\begin{eqnarray}
M_\delta \stackrel{d}{=} \sum_{j=1}^\infty W_j Y_j.\label{eq:DirichletMeanSum}
\end{eqnarray}
Here $W_1 = V_1$ and $W_j = V_j\prod_{i=1}^{j-1}(1-V_i), j\ge 2$ where $V_j, j\ge 1$ are i.i.d. variables equal in distribution to $\beta_{1,\delta}$ and $Y_j, j\ge 1$ are i.i.d. variables that have the same distribution as $Y.$ For some $\mathcal{N}\ge 1$, we can approximate $M_\delta$ by
\begin{eqnarray*}
M^{\mathcal{N}} = \sum_{j=1}^{\mathcal{N}} W_j Y_j + \left(1-\sum_{j=1}^{\mathcal{N}} W_j\right) Y_{\mathcal{N}+1}.
\end{eqnarray*}
Firstly, we consider the {\it fixed terms} truncation method that is similar as in \citet{IJ:2001}. It assumes that $\mathcal{N}$ is a fixed number and $Y$ can be either bounded or unbounded but have to satisfy ${\rm E}[Y]<\infty.$ By simple calculation, we get the error bound in $L^1$ sense as
$$
{\rm E}(|M_\delta - M^{\mathcal{N}}|) \le  {\rm E}[|Y|] \left(\frac{\delta}{\delta+1}\right)^{\mathcal{N}+1}.
$$
Secondly, we consider the {\it stopping time} approach of \cite{Guglielmi}. On the one hand, if $Y$ is bounded by a known constant $c_Y>0,$ the approach relies on detecting
$$
\mathcal{N} := \min_n\left\{n : c_Y\left(1-\sum_{j=1}^{n} W_j\right)<\epsilon , n=1,2,\ldots \right\}
$$
which is the first time that $M_\delta$ equals $M^{\mathcal{N}}$ up to a given machine precision $\varepsilon>0$ that is approximately 2.22e-016 for IEEE 754 standard \texttt{double}. We implement this number in the following numerical studies. The above stopping rule guarantees the following exact error bound rather than that in $L^1$ sense
$$
|M_\delta - M^{\mathcal{N}}|  \le \varepsilon.
$$
See also \citet{MT:1998} for related works. On the other hand, for unbounded cases with ${\rm E}[Y]<\infty$, we can easily extend the above stopping time idea as
$$
\mathcal{N} := \min_n\left\{n : {\rm E}[|Y|] \left(1-\sum_{j=1}^{n} W_j\right)<\epsilon , n=1,2,\ldots
\right\}
$$
which guarantees now an $L^1$ error bound as follows
$$
{\rm E}(|M_\delta - M^{\mathcal{N}}|) \le  \epsilon.
$$
See \citet{Murdoch:2000} for the modification of CFTP to deal with unbounded cases. In subsection
\ref{subsec:sim_GGC}, we compare the exact sampling method, ie the Double CFTP method, with approximation methods, showing their surprising high precision.
\begin{remark}
Based on the summation representation (\ref{eq:DirichletMeanSum}) of Dirichlet mean variables, the path simulation under OU-GGC models can be approximately handled as follows. By noticing that the {\em two elements} of (\ref{eq:TwoDtransPair}) and the possible {\em leverage increment} of (\ref{eq:Random_Pair_BNS_OU}) have $W$ (or equivalently $V$) variables in common, except that they have different $Y$ variables, we can first generate $\mathcal{N}$ copies of $V$ variables for both the two elements and the possible leverage increment, and then generate $Y$ variables respectively to get an approximate sample of the random pair (\ref{eq:TwoDtransPair}) and the possible leverage increment of (\ref{eq:Random_Pair_BNS_OU}).
\end{remark}
\begin{remark}
Note the infinite series representation of $M_{\delta}$ in~(\ref{eq:DirichletMeanSum}) is based on the sequence $(W_{i})$ having the size biased distribution of a sequence of probabilities with a Poisson Dirichlet law, otherwise known as a stick-breaking sequence; see \citet{IJ:2001} for pertinent references and background. The well-known fact that the tail of sum of this sequence, $\sum_{k=\mathcal{N}+1}^{\infty}W_{k},$ decreases at an exponential rate in $\mathcal{N}$ is key to the good performance of the approximation methods we use.
\end{remark}
\section{Simulation studies} \label{sec:Simulation}
All the simulation experiments are performed on a laptop PC with an Intel(R) Core(TM)2 Duo CPU P8700 2.53GHz and 1.89RAM. All programs are coded in C programming language with GNU scientific library (GSL) version 1.4 being used for generating random numbers from such standard distributions as Normal, Gamma, Beta and Uniform. See the reference manual of GSL for details about those standard random number generator algorithms and references therein. The C programs are compiled by Microsoft Visual C++ 6.0.

\subsection{The Double CFTP vs Approximation methods}\label{subsec:sim_GGC}
For $\delta\in\{0.1,0.2,0.5,1,2,5,10\},$ let $M$ be a Dirichlet mean variable with shape parameter $\delta$ and scale random variable given by $1-\exp(-\lambda U)$ according to our model setting. By simple calculation, the true mean and variance of $M$ are given by
$$
{\rm E}[M]= \exp(-1)/2 \mbox{ and } {\rm Var}[M]= (2/3\exp(-1)-5/12 \exp(-2)-1/6)/(\delta+1).
$$
Table \ref{table:cftp_app} displays experiment results of the Double CFTP and approximation algorithms with 10,000,000 trials. For approximation algorithms with fixed number of components, we consider $\mathcal{N}\in\{10,50, 100\}.$  All of the algorithms give reasonable results with small error from the true values. It implies that all considered approximation algorithms can be nice alternatives of the Double CFTP algorithm. On the other hand, the exact nature of the Double CFTP provides a reliable benchmark for comparison.

\begin{table}[tbhp]\renewcommand{\arraystretch}{1.2}\scriptsize
  \centering
    \vspace{0.10in}
    \caption{Simulation results of the Double CFTP and approximation methods
from $M$ with 10,000,000 trials.}
    \vspace{0.10in}
    \begin{center}
\begin{tabular}{*{3}{l}*{7}{r}}
\hline
\multicolumn{3}{c}{$\delta$}&0.1&0.2&0.5&1&2&5&10\\\hline\hline
\multicolumn{2}{c}{\multirow{2}*{}}&${\rm E}(M)$&\multicolumn{7}{c}{0.18393}\\
                   &&${\rm Var}(M)$&0.02018& 0.01850& 0.01480& 0.01110& 0.00740& 0.00370& 0.00202\\

\hline\hline

\multicolumn{2}{c}{\multirow{2}*{Double CFTP}}&$\hat {\rm E}(M)$&0.18399     &   0.18389     &   0.18390     &   0.18392     &   0.18398     &   0.18394     &   0.18395 \\
&&$\hat {\rm Var}(M)$&0.02019     &   0.01849     &   0.01479     &   0.01110     &   0.00740     &   0.00370     &   0.00202  \\
\hline

\multirow{9}*{App.}&\multirow{2}*{N=10}&$\hat {\rm E}(M)$&0.18397     &   0.18391     &   0.18396     &   0.18393     &   0.18391     &   0.18392     &   0.18394 \\
&&$\hat {\rm Var}(M)$&0.02017     &   0.01850     &   0.01480     &   0.01109     &   0.00742 &
0.00460 &   0.00592
 \\\cline{2-10}

&\multirow{2}*{N=50}&$\hat {\rm E}(M)$&0.18397     &   0.18391     &   0.18396     &   0.18393     &   0.18391     &   0.18392     &   0.18394 \\
&&$\hat {\rm Var}(M)$&0.02017     &   0.01850     &   0.01480     &   0.01109     &   0.00742     &
0.00460 &   0.00592\\\cline{2-10}

&\multirow{2}*{N=100}&$\hat {\rm E}(M)$&0.18397     &   0.18391     &   0.18396     &   0.18393     &   0.18391     &   0.18394     &   0.18394\\
&&$\hat {\rm Var}(M)$&0.02017     &   0.01850     &   0.01480     &   0.01109     &   0.00739 &
0.00370 &   0.00202\\\cline{2-10}

&\multirow{2}*{Stopping}&$\hat {\rm E}(M)$&0.18397     &   0.18391     &   0.18396     &   0.18393     &   0.18391     &   0.18394     &   0.18396 \\
&&$\hat {\rm Var}(M)$&0.02017     &   0.01850     &   0.01480     &   0.01109     &   0.00739     &   0.00370     &   0.00202\\
\hline\hline

\multicolumn{2}{c}{Double CFTP}&$\hat {\rm E}(\mathcal{S})$&76.14446    &   38.07715    & 15.22293 & 7.61515
& 15.23131    & 38.08296    &
76.14860\\
\multicolumn{2}{c}{App.-Stopping}&$\hat {\rm E}(\mathcal{N})$&4.74350     &   8.49265     &   19.76969    &   38.69694    &   77.41112    &   217.53297   &   1406.53346 \\
\hline
\end{tabular}
    \end{center}
    {\it Note.} The true values of the mean and variance of $M$ are ${\rm E}(M)= 0.18393$ and ${\rm Var}(M)= 0.02220/(\delta+1)$ respectively. Here $\hat {\rm E}(M)$ is a mean estimate, $\hat {\rm Var}(M)$ is a variance estimate, $\hat {\rm E}(\mathcal{S})$ is an estimate of ${\rm E}(\mathcal{S})$ and $\hat {\rm E}(\mathcal{N})$ is an estimate of ${\rm E}(\mathcal{N})$.
\label{table:cftp_app}
\end{table}

\begin{figure}
\begin{center}
    \includegraphics*[width=4.0in, height=4.0in]{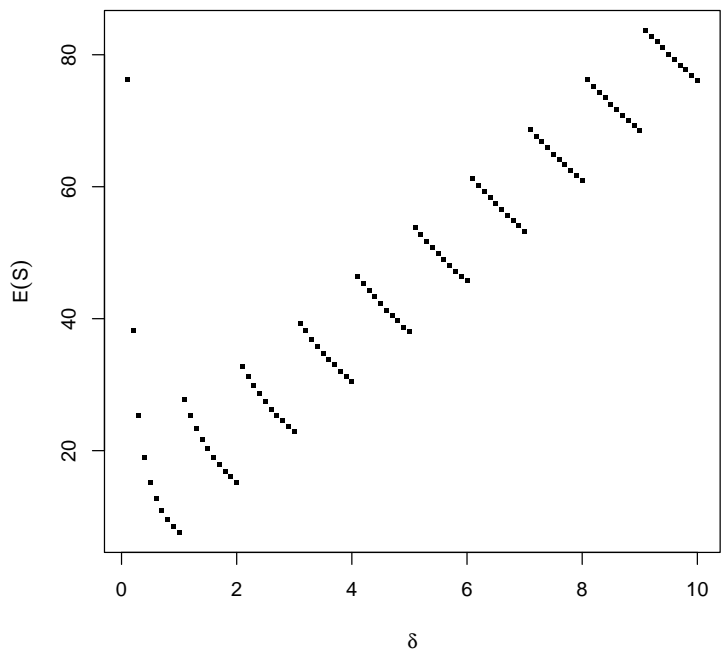}
    \end{center}
    \caption{Plot of the average numbers of total stack size in the Double CFTP
    algorithm for ($\delta=0.1,~0.2,~0.3,~\ldots,~9.9~{\rm and}~10.$) with 10,000,000 trials.}
\label{fig:ES}
\end{figure}

Let $\mathcal{S}$ be the stack size of the Double CFTP algorithm, see Appendix B. Computation times and consumption of memory heavily depend on ${\rm E}(\mathcal{S})$ for the Double CFTP and ${\rm E}(\mathcal{N})$ for a truncating algorithm with stopping time. The main problem with Double CFTP algorithms is that ${\rm E}(\mathcal{S})$ is not uniformly bounded. It is especially bad when $\delta$ is very small or very large. Figure \ref{fig:ES} shows estimates of ${\rm E}(\mathcal{S})$ for $\delta=0.1,0.2,0.3,\ldots,9.8,9.9,10.0$. The minimum is achieved when $\delta=1.$ For a truncating algorithm with a stopping time, estimates of ${\rm E}(\mathcal{N})$ keep increasing as $\delta$ is increasing, see also \citet{DJ:2010} for more discussions.

\subsection{From BNS OU-Gamma to GL-BNS OU-GGC}\label{subsec:sim_GAMMA_GGC}
\begin{figure}
    \includegraphics*[width=3.0in, height=3.0in]{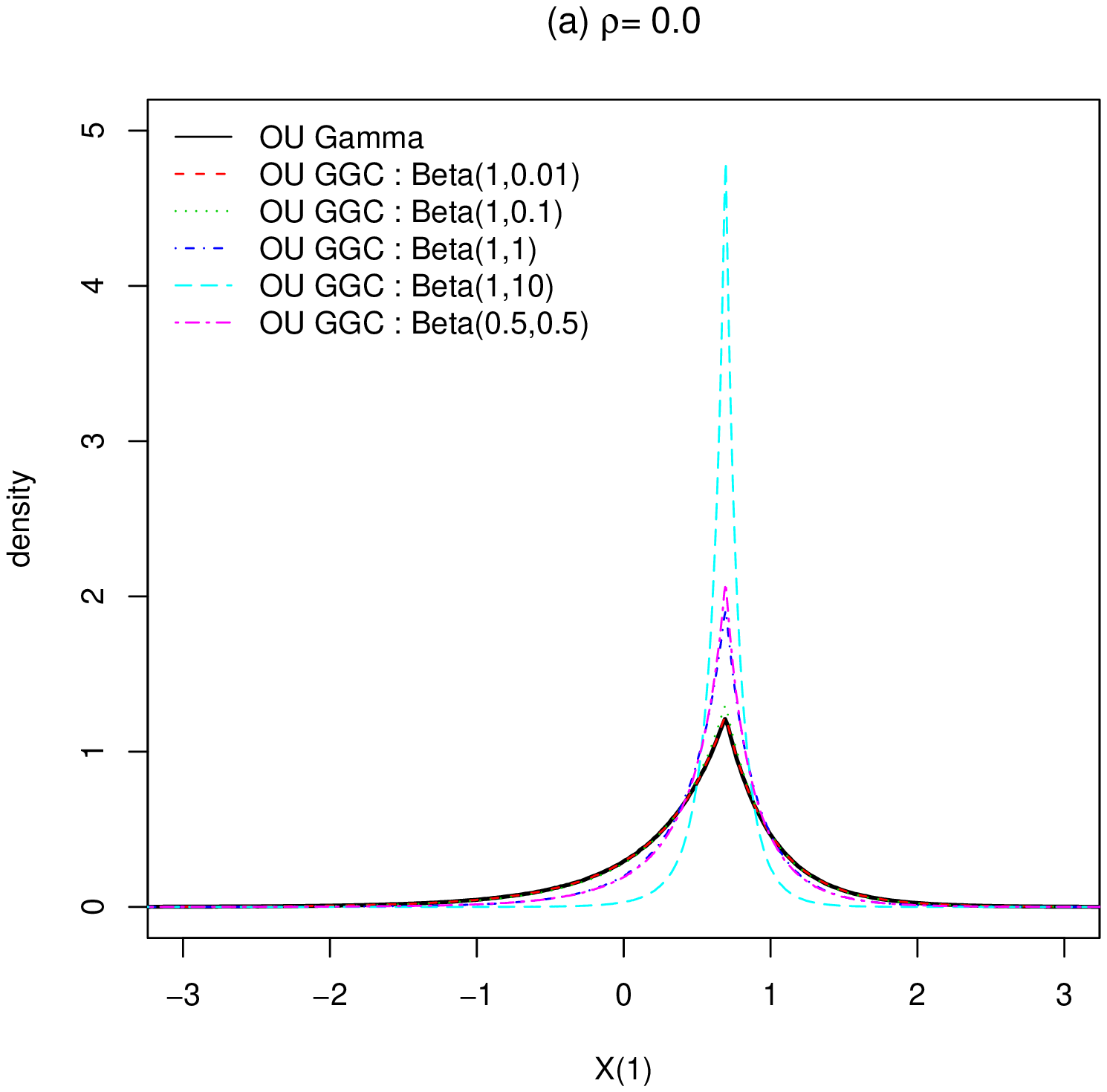}
    \includegraphics*[width=3.0in, height=3.0in]{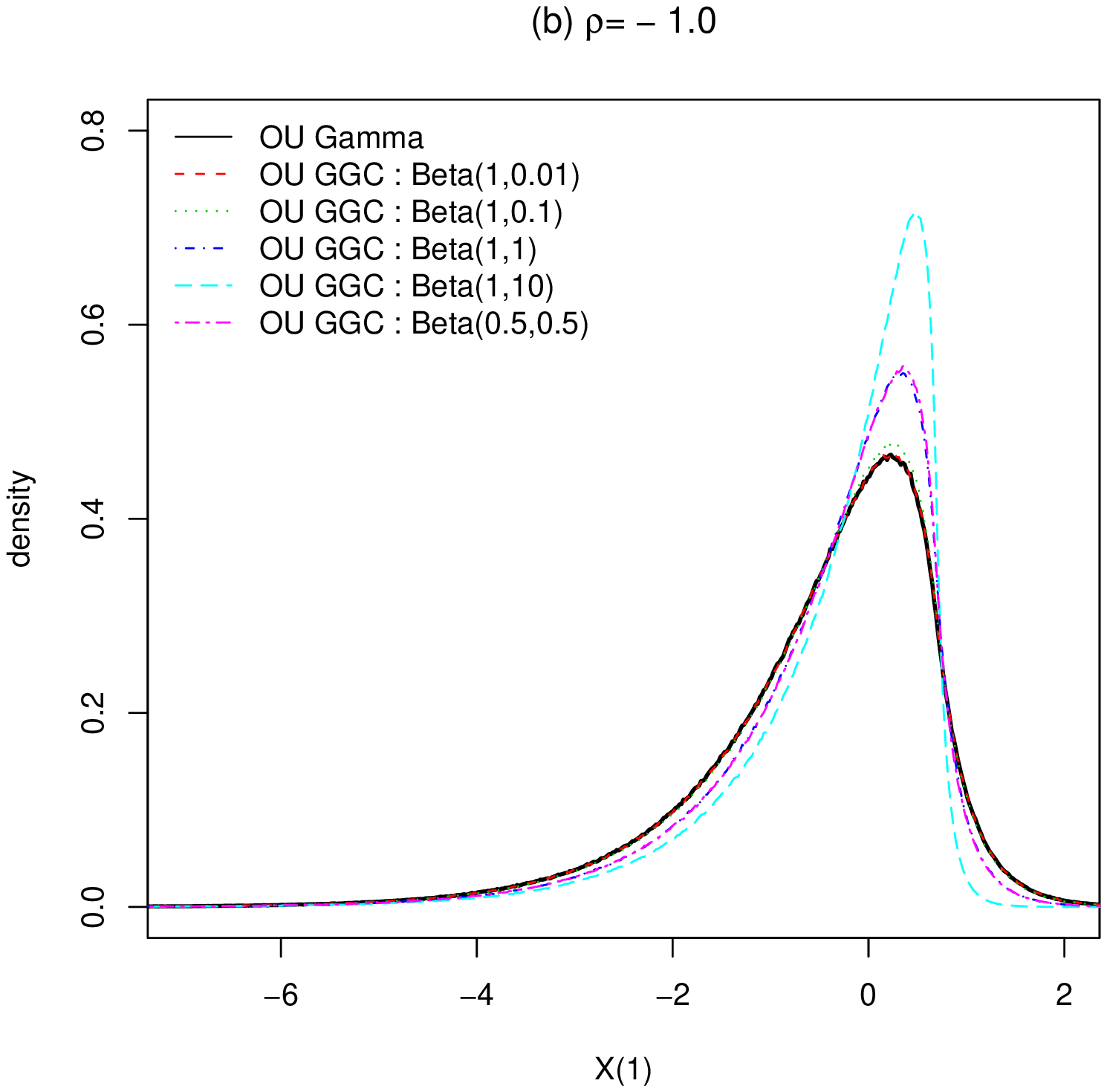}
    \caption{Estimated densities of $X(1)$ over 10,000,000 trials. $\theta=c=\lambda=1$, $v(0)=0$ and $r=q= 0.$}
\label{fig:gamma_ggc}
\end{figure}

In this section we compare the BNS OU-Gamma and GL-BNS OU-GGC models in terms of the flexibility of generating different distributions for the log return, eg $X(1)$, when
$\rho = 0$ and $\rho = -1.$ For simplicity, we assume that $\theta=c=\lambda=1,$ $v(0)=0,$ $X(0)=0$ and $r=q=0.$ Under these assumptions, $X(1)$ is simplified to
$$
X(1)|\gamma(1),\tau(0,1)\sim {\rm Normal}(\mu(0,1),\tau(0,1)),
$$
where now $\mu(0,1) = \log(1-\rho)-\tau(0,1)/2 + \rho \gamma(1)$ and $\tau(0,1) =
\int_0^1(1-\exp(-(1-s)))dZ( s).$ Here $Z$ is either a Gamma process with shape $1$ and scale $1$ for the BNS OU-Gamma model or a GGC$(1,R)$ subordinator with $R\sim{\rm Beta}(a,b),(a,b)\in \{(1,0.01),(1,0.1),(1,1),(1,10),(0.5,0.5)\}$ for GL-BNS OU-GGC model. Figure \ref{fig:gamma_ggc} displays the estimated densities of $X(1)$ over 10,000,000 trials. Table \ref{table:gamma_ggc} gives the mean, standard deviation, skewness and kurtosis of $X(1)$ over 10,000,000 trials. These simulation results show that the GL-BNS OU-GGC model is a reasonable extension to the BNS OU-Gamma model in terms of distributional flexibility.

\begin{table}[tbhp]\renewcommand{\arraystretch}{1.2}\footnotesize
  \centering
    \vspace{0.10in}
    \caption{Estimated mean, standard deviation(s.d.), skewness(skew.) and kurtosis(kurt.) of $X(1)$ over 10,000,000 trials.}
    \vspace{0.10in}
    \begin{center}
\begin{tabular}{ccrrrrrr}\hline
\multicolumn{2}{c}{}&BNS&\multicolumn{5}{c}{GL-BNS OU-GGC}\\\cline{4-8}
\multicolumn{2}{c}{}&OU-Gamma&Beta(1,0.01)&Beta(1,0.1)&Beta(1,1)&Beta(1,10)&Beta(0.5,0.5)\\

\hline\hline \multirow{4}{*}{$\rho=0$}
&mean & 0.5090   &   0.5111   &    0.5256   &    0.6011   &    0.6765   &   0.6014\\
&s.d. & 0.6404   &   0.6367   &    0.6091   &    0.4447   &    0.1846   &   0.4465\\
&skew.&-1.0415   &  -1.0423   &   -1.0463   &   -1.0125   &   -0.6272   &  -1.1309\\
&kurt.& 7.6148   &   7.6475   &    7.8050   &    8.9646   &   10.4274   &   9.7664\\

\hline\hline\multirow{4}{*}{$\rho=-1$}
&mean &-0.4912   &  -0.4883   &   -0.4745   &   -0.3992   &   -0.3229   &  -0.3983\\
&s.d. & 1.1876   &   1.1849   &    1.1709   &    1.0945   &    1.0160   &   1.0955\\
&skew.&-1.3571   &  -1.3576   &   -1.3925   &   -1.5905   &   -1.9043   &  -1.6040\\
&kurt.& 6.4008   &   6.3918   &    6.5259   &    7.3110   &    8.6068   &   7.4114\\
\hline
\end{tabular}
    \end{center}
    {\it Note.} $\theta=c=\lambda=1$, $v(0)=0$ and $r=q= 0.$
\label{table:gamma_ggc}
\end{table}

\section{Numerical results on simulation pricing}\label{sec:Numerical_Result}
In this section, we investigate the performance of our proposed models in terms of pricing both the European call options (path independent) and the forward start options (path dependent). The study consists of two parts. First, we calibrate our proposed models to a real option data set that was used by many authors in the literature. Based on the parameter estimation from the calibration procedure, we then demonstrate the capability of those proposed models in pricing some of the path dependent options, eg the forward start options.

We call the price generated by directly taking the Monte Carlo average of payoffs calculated over sample paths of the asset price plain simulation price (PSP), see section \ref{sec:simulatingPrice}. In order to improve the efficiency of the exact simulation pricing method, a conditional BS representation of the derivative price can be applied to reduce the Monte Carlo error of the price estimator. See (\cite{HW:1987} and \cite{Willard:2003}) for the argument of this representation under SV models. The price produced by this latter method is called formula simulation price (FSP). Next we explain how the conditional Monte Carlo method works. We follow the sections 5 and 7 of \cite{BK:2006} and subsection 4.2 of \cite{NV:2003}.

\subsection*{European Call options}
The price of a European call option ${\rm E}[e^{-r(T-t)}(S(T)-K)^+|S(t),v(t)]$ can be equivalently represented under BNS OU SV models as
$$
{\rm E}[BS(S_0,\sigma_0|Z(\cdot))],
$$
where $BS(\cdot,\cdot)$ denotes the usual BS price formula, $S_0$ and $\sigma_0$ are the corresponding spot price and volatility and the notation emphasizes that the BS price is conditional on a path of the BDLP $Z(t).$ Under the BNS OU SV model (\ref{eq:BNSOU}), we have that $S_0 = S(t)\exp(-\lambda\kappa(\rho)(T-t) + \rho \int_t^TdZ(\lambda s))$ and $\sigma_0 = \sqrt{\tau(t,T)/(T-t)}.$ Therefore, it is not necessary to condition on the entire path of the BDLP. The call price can just be rewritten as
$$
{\rm E}\left[BS\left(S_0,\sigma_0|\int_t^TdZ(\lambda s),\int_t^T(1-e^{-\lambda(T-s)})dZ(\lambda s)\right)\right].
$$
Under the BNS OU-Gamma model, conditional on $(S(t),v(t)),$ we can produce exactly $B$ i.i.d. copies of the random pair (\ref{eq:Random_Pair_BNS_GAMMA}), and hence $B$ i.i.d. copies of the BS price inside the expectation. The FSP is therefore written as $(1/B)\sum_{b=1}^BBS^b$, in contrast to the PSP given by $(1/B)\sum_{b=1}^Be^{-r(T-t)}(S^b(T)-K)^+.$ We denote
$$
C_E(S(t),K,T-t,v(t)):=BS\left(S_0,\sigma_0|\int_t^TdZ(\lambda s),\int_t^T(1-e^{-\lambda(T-s)})dZ(\lambda s)\right)
$$
for later use.

\subsection*{Forward Start Options}
A similar conditional argument can be applied to pricing a forward start option with fair price $e^{-rT_2}{\rm E}\left[(S(T_2) - k S(T_1))^+ \right].$ By the linear homogeneity of the European call price w.r.t. the spot price and the strike, we can rewrite the forward start option price as
$$
{\rm E}\left[e^{-rT_1}S(T_1)C_E(1,k,T_2-T_1,v(T_1))\right],
$$
where $C_E(1,k,T_2-T_1,v(T_1))$ is the conditional BS price as defined in the previous European call option case. Therefore, under the BNS OU-Gamma model, conditional on $(S(0),v(0)),$ we can sample exactly $B$ i.i.d. copies of random pairs (\ref{eq:Random_Pair_BNS_GAMMA}) and (\ref{eq:TwoDtransPair_GAMMA}), and hence $B$ i.i.d. copies of $S(T_1)$ and also the BS price inside the expectation. The FSP is thus given by $(e^{-rT_1}/B)\sum_{b=1}^B
S^b(T_1)C^{b}_E,$ in contrast to the PSP given by $(e^{-rT_2}/B) \sum_{b=1}^B (S^b(T_2) - k S^b(T_1))^+$. See section 7 of \cite{BK:2006} for more details and \citet{KN:2005} for relevant works.

\subsection{Model calibration}\label{subsec:calibration}
The data set has 87 call option prices on the S\&P 500 Index at the close of the
market on 2 November 1993, see (\cite{AL:1998} and \cite{AL:2000}) for a detailed description of the data set. On that day, the S\&P 500 Index closed at 468.44. The risk-free rate $r$ is $3.19\%$ and the dividend yield $q$ is assumed to be zero. This data set has been analyzed by many authors including (\citet{DSP:2000} and \citet{NV:2003}).

We compared the BNS OU-Gamma and GL-BNS OU-GGC models both with superposition $l=1,2.$ For the GL-BNS OU-GGC model, we used $R$ such that $R/c \sim {\rm Beta}(\alpha,\beta),c>0,\alpha>0~{\rm and}~\beta>0.$ For models with superposition $l=2$, we give a constraint on the parameter set that $\lambda_1 \ge \lambda_2$ to avoid the
label-switching problem, ie the unidentifiability of the permutation of $\lambda_i$'s, see \citet{DR:1994} for a label-switching problem. Typically, calibration (fit of model to the market data) is a model estimation procedure under the risk neutral measure, which is done by minimizing the mean-square error
$$
\mbox{MSE} = \sum_{{\rm options}} \frac{(\mbox{market price}-\mbox{model price})^2}{\mbox{number of
options}},
$$
where model prices are the simulation estimates of the expected discounted payoffs (the true model prices). Therefore, the MSE that we report here involves a Monte Carlo error, ie the variance of the simulation estimator of the true model price. FSP is used to reduce this Monte Carlo error. For the calibration purpose, the Double CFTP perfect sampler is available for all the above considered models. Usually a grid search method is used to find parameter estimates. However, a grid search over more than 5 dimensional parameter space is almost impossible. Instead of a grid search, we used the simplex algorithm for multidimensional minimization problems (\cite{NM:1965}) that is implemented in a GSL version 1.4. It is one of those methods that do not require derivatives and are robust. We use 100,000 trials, ie $B=100,000$, to get model prices. This number turns out to be large enough so that the MSE reported here is precise up to at least three decimal places. Table \ref{table:calib} displays the risk-neutral calibrated parameters and MSEs. There is notable improvement of fit with introducing the GGC extension while little improvement can be made by including superposition. In terms of MSE, the OU-Gamma model outperforms the Heston model, by noticing that OU-Gamma model has an MSE of $0.00870$ (see second column of Table \ref{table:calib}) while the MSE of Heston model is $0.0124$ (this can be found from the first column of Table $I$ of \cite{DSP:2000}) and the MSE of OU-Gamma model is significantly smaller. As can be seen from the Table 5.1 of \citet{NV:2003}, BNS models with IG and Gamma marginals give MSEs $0.01686$ and $0.01244$ respectively, hence OU-Gamma model outperforms these two models as well.

\begin{table}[tbhp]\renewcommand{\arraystretch}{1.2}\footnotesize
  \centering
    \vspace{0.10in}
    \caption{Fitted parameter values for 4 models.}
    \vspace{0.10in}
    \begin{center}
\begin{tabular}{crrrr}\hline
&\multicolumn{2}{c}{BNS OU-Gamma}&\multicolumn{2}{c}{GL-BNS OU-GGC}\\\cline{2-5}
super-pos.&\multicolumn{1}{c}{$l=1$}&\multicolumn{1}{c}{$l=2$}&\multicolumn{1}{c}{$l=1$}&\multicolumn{1}{c}{$l=2$}\\\hline\hline

$\rho$          &--~4.88115 &--~4.87261 &  --~0.04455 &--~0.04434 \\
$\theta$        &   0.81303 &   0.79608 &     0.80124 &   0.79536 \\
$c$             &   0.00981 &   0.00989 &     0.00954 &   0.00971 \\
$\alpha$        &           &           &     3.61908 &   3.68861 \\
$\beta$         &           &           &     0.10414 &   0.11126 \\
\hline
$v^{(1)}(0)$    &   0.00437 &   0.00418 &     0.00414 &   0.00407 \\
$\lambda_1$     &   2.24323 &   2.27276 &     2.68545 &   2.69492 \\
$v^{(2)}(0)$    &           &   0.00006 &             &   0.00003 \\
$\lambda_2$     &           &   0.02755 &             &   0.01442 \\

\hline\hline

MSE             &   0.00870 &   0.00842 &     0.00778 &   0.00711 \\
\hline
\end{tabular}
    \end{center}
\label{table:calib}
\end{table}

\subsection{Simulating path dependent option prices}
In Table \ref{table:FSO}, we also give the simulation pricing results for forward start options based on parameters estimated from the previous calibrations. For BNS OU-Gamma model, we applied both the exact sampling method and approximation methods for comparison. Estimated prices are similar for both methods. The time needed for the exact method is more than the approximation method under the BNS OU-Gamma model with superposition and the set of estimated parameters. For the GL-BNS OU-GGC model, we can only apply approximation methods. It also gives similar estimated values. The Monte Carlo errors of FSPs are significantly smaller than that of PSPs. It is not surprising that a large amount of time is required in the exact path simulation of OU-Gamma model with superposition $l=2,$ since, as we mentioned earlier in section \ref{subsec:Sup_OU_gamma}, the two independent OU processes have to be treated separately in order to generate a price path. However, according to our fitted result in Table \ref{table:calib}, the second OU process results in simulating a Dirichlet Mean variable with shape parameter $\delta=\theta\lambda_2\approx0.02,$ which is time consuming as expected, see section \ref{subsec:sim_GGC} and in particular Figure \ref{fig:ES}. Nonetheless, this is not a problem for the path independent case as in model calibration, where one can always treat the simulation in an aggregate fashion as indicated by Proposition \ref{prop:sup} of subsection \ref{subsec:GL_sup}. The related Dirichlet Mean variable then has shape parameter $\delta=\theta(\lambda_1+\lambda_2)$ with moderate value.

Although we illustrate here the exact simulation pricing of forward start options, other more heavily path dependent derivatives, eg Bermudan options, can also be priced by the proposed exact simulation method as long as the derivatives' payoff function depends on discretely spaced asset prices at a finite number of times. The computation time is linear in the number of time points when they are equally spaced.

\begin{table}[tbhp]\renewcommand{\arraystretch}{1.2}\footnotesize
  \centering
    \vspace{0.10in}
    \caption{Simulation estimates for a forward start option under different estimated models.}
    \vspace{0.10in}
    \begin{center}
\begin{tabular}{ccccrrrrrr}\hline
\# of trials&\multicolumn{3}{c}{}&PSP&s.e.&c.t.&FSP&s.e.&c.t.\\\hline\hline

\multirow{6}{*}{10,000}&\multirow{4}{*}{BNS OU-Gamma}& \multirow{2}{*}{Exact}
  &$l=1$&5.95545&0.06524&2  &5.97314&0.02963&2\\
&&&$l=2$&5.98338&0.06593&114&6.00002&0.02973&113\\\cline{3-10}

&&\multirow{2}{*}{App.}
  &$l=1$&5.98782&0.06433&2&5.99087&0.02904&2\\
&&&$l=2$&6.04260&0.06412&3&5.98026&0.02934&3\\\cline{2-10}

&\multirow{2}{*}{GL-BNS OU-GGC}& \multirow{2}{*}{App.}
  &$l=1$&6.00127&0.06489&2&6.06638&0.02890&2\\
&&&$l=2$&5.96249&0.06314&3&5.96251&0.02895&3\\\hline\hline

\multirow{6}{*}{100,000}&\multirow{4}{*}{BNS OU-Gamma}& \multirow{2}{*}{Exact}
  &$l=1$&5.98164&0.02127&15 &5.98314&0.00941&16\\
&&&$l=2$&5.99319&0.02145&1131&6.00002&0.00938&1123\\\cline{3-10}

&&\multirow{2}{*}{App.}
  &$l=1$&6.00002&0.02125&16&5.99100&0.00940&15\\
&&&$l=2$&6.00211&0.02201&29&6.00006&0.00951&28\\\cline{2-10}

&\multirow{2}{*}{GL-BNS OU-GGC}& \multirow{2}{*}{App.}
  &$l=1$&5.99176&0.02103&17&5.98917&0.00939&17\\
&&&$l=2$&5.98142&0.02192&29&5.99251&0.00941&29\\\hline
\end{tabular}
    \end{center}
    {\it Note.} Fitted model parameters are shown in table \ref{table:calib} and $S(0) = 100$, $k=1$, $T_1=1$ year, $T_2=2$ years. 100 components are used for approximation methods. Here s.e. is standard error and c.t. is computing time (in seconds).
\label{table:FSO}
\end{table}
\section{Conclusions}
In this paper, we devise the exact path simulation method for the nontrivial BNS type OU-Gamma SV models that parallels the result of \cite{BK:2006} for the popular Heston model. Such extensions as superposition of independent OU volatility processes and OU-GGC models to the baseline OU-Gamma models are also studied. The exact simulation technique still applies to pricing path independent derivatives under the larger class of OU-GGC models by introducing the gamma leveraging concept, while approximation method is needed when considering pricing path dependent derivatives. The high precision of the two proposed approximation methods is confirmed by a comparison to the exact Double CFTP method in the numerical study. We then calibrated the proposed OU-Gamma model among others to a well known European option data set that has been studied by many authors in the literature, showing its better performance than the Heston model and two other BNS type models. The models' capabilities of pricing path dependent options are also illustrated by numerical tests, showing their potential in practical applications.


\section*{Appendix A: Independence properties of GGC random variables }\label{app:indep}
Let $N$ denote a Poisson random measure on $\mathbb{R}_+^3$ with mean measure
$$
E[N(ds,dr,dy)] = \theta s^{-1} \exp(-s) dsF_R(dr) dy
$$
where $\theta>0$ and $F_R$ is a probability measure. A generalized Gamma convolution subordinator, GGC$(\theta,R)$,  is defined as
\begin{eqnarray*}
Z(t) := \int_0^t\int_0^\infty\int_0^\infty rs N(ds,dr,dy)= \int_0^t\int_0^\infty
r\gamma(dr,dy). \label{eq:BDLP}
\end{eqnarray*}
As a special case, if $R$ is a constant $c$, it is just a Gamma process with shape parameter $\theta>0$ and scale parameter $c>0.$ Here
$$
\gamma(dr,dy) := \int_0^\infty s N(ds,dr,dy)
$$ is a Gamma completely random measure on $\mathbb{R}_+^2$ with mean measure $E[\gamma(dr,dy)] = \theta F_R(dr)
dy := \eta(dr,dy).$ That is, for any disjoint subsets of $\mathbb{R}_+^2$ $A$ and $B$,
$\gamma(A)$ and $\gamma(B)$ are independent Gamma$(\eta(A))$ and Gamma$(\eta(B))$ random
variables respectively, where
$$
(\gamma(A),\gamma(B)) = \left(\int_A \gamma(dr,dy), \int_B \gamma(dr,dy)\right).
$$
Additionally,
$$
\gamma(t) := \int_0^t \int_0^\infty \gamma(dr,dy)
$$
is a Gamma subordinator that is the extracted Gamma component of $Z(t)$. For $r\in[0,\infty)$ and $y\in[0,t]$,
$$
\left\{ \frac{\gamma(dr,dy)}{\gamma(t)}\right\}
$$
is a random probability measure which is a Dirichlet process and is independent of $\left\{
\gamma(s):s\ge t\right\}$. It follows that for any non-negative function $k$,
$$
\int_0^t k(y) Z(dy) = \int_0^t \int_0^\infty k(y) r \gamma(dr,dy) = \gamma(t) M_t.
$$
Here $M_{t}$ is a Dirichlet mean variable that is independent of  $\gamma(t)$ and satisfies
\begin{eqnarray*}
M_{t} \stackrel{d}{=} V Y+(1-V)M_{t}\label{eq:MVY}
\end{eqnarray*}
where $Y=k(Ut)R$, $V\sim{\rm Beta}(1,\theta t)$ and $U\sim{\rm Uniform}[0,1]$. ${\rm
Uniform}[a,b]$ denotes a uniform distribution on $[a,b]$ and ${\rm Beta}(a,b)$ denotes a beta distribution with mean $a/(a+b)$ and variance $ab/((a+b)^2 (a+b+1))$. Three random variables on the right-hand side, ie $Y$, $V$ and $M_{t}$, are mutually independent. While $\gamma(t)$ and $\int_0^t k(y) Z(dy)$ are clearly dependent, the distribution of the pair can be expressed as
$$
\left(\gamma(t), \int_0^t k(y) Z(dy)\right) \stackrel{d}{=} (\gamma(t), \gamma(t) M_{t}).
$$
See for instance \cite{JRY:2008} and \cite{James:2010} among others for references.

\section*{Appendix B: A compact algorithm of the Double CFTP}\label{app:CFTP}
The Double CFTP method of \cite{DJ:2010} is designed for exact sampling a random variable defined through the following stochastic equation or Markovian map
\begin{align}
M\overset{d}=VY+(1-V)M,\notag
\end{align}
where all variables on the right-hand side are independent of one
another and $M$ on both sides have the same distribution. We assume that the density $h$ of $V$ is bounded from below on $[0,1]$ by a constant $c_h$ and $0<Y\leq c_Y<\infty.$ In the following algorithm, see \cite{JZ:2010}, $(U_i)_{i\geq1}$ are ${\rm Uniform}[0,1]$ variables and $Y$ and $Y'$ have the same distributions.

\begin{quote}
\begin{description}
\item[Backward phase.] For $i=-1,-2,\ldots$: keep generating $(U_i,Y_i,Y'_i)$ and
storing $(Y_i,Y'_i)$ until $U_\mathbb{T}\leq
|Y_\mathbb{T}-Y'_\mathbb{T}|c_h/(2c_Y).$ Keep $\mathbb{T}.$
\item[Set starting point.] Set $M=Y_\mathbb{T}\wedge
Y'_\mathbb{T}+2c_YU_\mathbb{T}/c_h.$
\item[Forward phase.] For $i=\mathbb{T}+1,\mathbb{T}+2,\ldots,-1$: given $(Y_i,Y'_i,M)$ previously stored, do the following step: generate $U'\sim {\rm Uniform}[0,1]$, $\xi_{1/2}\sim{\rm Bernoulli}(1/2)$ and $V$, and construct $X=(1-V)M+VY_i\xi_{1/2}+VY'_i(1-\xi_{1/2}).$ Repeat this step until: $$U'[h(\frac{X-M}{Y_i-M})\frac{1}{|Y_i-M|}
    +h(\frac{X-M}{Y'_i-M})\frac{1}{|Y'_i-M|}]> c_h/c_Y$$ or $X< Y_i\wedge Y'_i$ or $X> Y_i\vee Y'_i$. Then set $M=X.$
\item[Output.] Return $M$.
\end{description}
\end{quote}
As a special case, $M$ is a Dirichlet mean variable $M_{\delta}$
when $V$ is a ${\rm Beta}(1,\delta)$ variable. Particularly in this paper, the Dirichlet mean variables under consideration are those with $V\sim{\rm Beta}(1,\delta)$ for $0<\delta\leq 1$, whose density is $h(x)=\delta(1-x)^{\delta-1}\leq\delta$ for all $x\in[0,1]$, and $Y$ being $R,$ $R\exp(-\lambda U(T-t))$ or $R(1-\exp(-\lambda U(T-t)))$ (see section 5), which are bounded by $c$, $c$ and $c(1-\exp(-\lambda (T-t)))$ respectively if $R\leq c$. The above algorithm is ready-to-use for simulating relevant Dirichlet mean random numbers by simply setting $c_h=\delta$ and $c_Y=c~{\rm or}~c(1-\exp(-\lambda (T-t))).$

\section*{Acknowledgments}
We would like to thank Prof. Yacine A\"it-Sahalia for providing us the option data set. Dohyun Kim's research is supported by the Korean Research Foundation Grant funded by the Korean Government (KRF-2008-314-C0046).





\bibliographystyle{nonumber}

\end{document}